\documentclass[12pt,preprint]{aastex}
\usepackage{apjfonts}
\usepackage{natbib}
\usepackage{lscape}

\shorttitle{TP-AGB Luminosity Bound}
\shortauthors{Bird \& Pinsonneault}

\newcommand{\eg}{{\rm e.g.}}
\newcommand{\cf}{{\rm cf.}}
\newcommand{\ie}{{\rm i.e.}}

\newcommand{\teff}{\ensuremath{T_{\mathrm{eff}}}}
\newcommand{\mf}{\ensuremath{M_\mathrm{f}}}
\newcommand{\mi}{\ensuremath{M_\mathrm{i}}}
\newcommand{\mtp}{\ensuremath{M_{\mathrm{c,1TP}}}}
\newcommand{\tcool}{\ensuremath{\tau_{\mathrm{cool}}}}
\newcommand{\tclus}{\ensuremath{\tau_{\mathrm{clus}}}}
\newcommand{\tprog}{\ensuremath{\tau_{\mathrm{prog}}}}
\newcommand{\feh}{\ensuremath{[\mathrm{Fe/H}]}}
\newcommand{\delmc}{\ensuremath{\Delta M_\mathrm{c}}}
\newcommand{\logg}{\ensuremath{\log\ g}}
\newcommand{\dm}{\ensuremath{(m-M)}}
\newcommand{\msol}{\ensuremath{M_{\odot}}}
\newcommand{\fecore}{\ensuremath{f_{\mathrm{E, \Delta M_c}}}}
\newcommand{\fhe}{\ensuremath{F_{\mathrm{TP-AGB, He}}}}
\newcommand{\lmin}{\ensuremath{\mathcal{L}_{\mathrm{min}}}}
\newcommand{\lmod}{\ensuremath{\int L_{\mathrm{M}_\mathrm{i}, \feh}(t)dt}}
\newcommand{\fctp}{\ensuremath{f_{\mathrm{M_c,TP-AGB}}}}

\begin{document}

\title{A Bound on the Light Emitted During the TP-AGB Phase}

\author{Jonathan C. Bird, Marc H. Pinsonneault}

\affil{Department of Astronomy, The Ohio State University, 140 West 18th
Avenue, Columbus, OH 43210, bird@astronomy.ohio-state.edu,
pinsono@astronomy.ohio-state.edu}

\begin{abstract}

The integrated luminosity of the TP-AGB phase is a major uncertainty in stellar population synthesis models. We use the white dwarf initial final mass relation and stellar interiors models to demonstrate that a significant fraction of the core mass growth for intermediate ($1.5<\msol<6$) mass stars takes place during the TP-AGB phase. We find evidence that the peak fractional core mass contribution for TP-AGB stars is $\sim20\%$ and occurs for stars between $2\ \msol$ and $3.5\ \msol$. Using a simple fuel consumption argument we couple this core mass increase to a lower limit on the TP-AGB phase energy output. Roughly half of the energy released in models of TP-AGB stars can be directly accounted for by this core growth; while the remainder is predominantly the stellar yield of He. A robust measurement of the emitted light in this phase will therefore set strong constraints on helium enrichment from TP-AGB stars, and we estimate the yields predicted by current models as a function of initial mass. Implications for stellar population studies and prospects for improvements are discussed.

\end{abstract}

\section{Introduction}\label{sec:intro}

Stars can experience a core-collapse supernova only if they are born with a mass much higher than the Sun, even though a chemically evolved core of order only one solar mass is required to ignite the advanced burning stages.  The culprit is mass loss severe enough to strip off the stellar envelope before a sufficiently massive processed core develops.  Although there is some mass loss even in shell hydrogen burning giants, the vast majority in intermediate mass stars occurs in the presence of thermal pulsations involving interactions between hydrogen and helium burning shells; we refer to this as the thermally pulsing AGB (TP-AGB) phase.  Our understanding of even the basic properties of the TP-AGB phase, such as the lifetime or light emitted, is limited because their severe mass loss is difficult to constrain observationally or predict theoretically.
 
The uncertainties in TP-AGB evolution have profound consequences for stellar population studies.  Stellar interiors models can accurately predict evolutionary properties up to the onset of the TP-AGB phase, and such models have been extensively used to reconstruct star formation histories in both resolved and unresolved populations.    Models of the TP-AGB phase, however, are extremely dependent on the assumptions related to mass loss and their predictive power is thus limited.  There has been intriguing work suggesting that the fraction of light emitted by TP-AGB stars in LMC star clusters is very high, of order 40 \% \citep{Persson83}. Such a substantial flux, largely redistributed to the far-IR by dust, can be especially important for interpreting intermediate redshift galaxy properties \citep{Conroy09}. There has been a strong emphasis on updating population synthesis to include the TP-AGB phase \citep[\eg][]{Bruzual03, Maraston05}. \citet{Maraston06} demonstrated that modeling of the TP-AGB phase has become a defining characteristic of different stellar population synthesis (SPS) codes while \citet{Conroy09} identified the substantial uncertainties in TP-AGB properties as a major component of the error budget for galaxy evolution models.

In this paper we employ a fuel consumption argument to set a firm lower bound on the fraction of light emitted during the TP-AGB phase.  We use stellar interiors models to set the core mass at the onset of the TP-AGB phase, and demonstrate that it is surprisingly insensitive to the choice of input physics.  The nuclear processed core then grows until the envelope is expelled, at which point the final white dwarf mass is set.  The white dwarf initial-final mass relationship (IFMR) can in turn be inferred from open clusters.  We use the difference between the final and starting masses in the TP-AGB phase as a bound on the fuel consumed, and thus the emitted light.  Although it is common for investigators to compare their models to initial-final mass relationships, and fuel consumption arguments have been used to check population synthesis models, the quantitative bounds from the IFMR are typically not used in population synthesis calculations.  We argue that the additional empirical information encoded there permits a narrower set of possibilities than considered by \citet{Conroy09}. Processed fuel (especially helium) can be ejected in winds, so the white dwarf data formally sets only a lower bound on the emitted light.  This raises interesting links between the light emitted in this phase and chemical evolution studies, which we discuss in our conclusions. The plan of our paper is straightforward.  Our sample and methods are discussed in Section 2, our results are presented in Section 3, and we discuss their broader implications in Section 4.

\section{Sample and Methods}\label{sec:methods}

Stellar evolution theory makes a robust prediction for the core mass at the onset of the TP-AGB phase as a function of composition and initial mass. Open clusters provide a laboratory where we can constrain the initial mass given a sample of known stellar remnants. Through fuel consumption arguments, the difference between the final core mass (\mf), set by the mass of the white dwarf, and the core mass at the onset of the TP-AGB phase (\mtp) is a lower limit on the total light emitted from stars during their TP-AGB phase as a function of initial mass. We therefore begin by discussing the white dwarf IFMR, which provides the largest and most accurate sample of initial-final mass pairs. We then address the theoretical interior models and their uncertainties, and close with how we relate fuel consumption to energy output.

Our sample consists of 48 white dwarfs (WDs) in 9 open clusters. The primary reference for the majority of our (\mi - \mf) data is the compilation of \citet{Salaris09}; hereafter S09. S09 provide an extensive investigation of the systematic errors involved in determining the IFMR and calculate initial and final masses self-consistently across their entire sample, \ie, they use the same set of physics to compute the WD masses, cluster ages, and initial masses. We limit our analysis to clusters appearing in the S09 study. We include initial and final mass pairs from nine of the ten open clusters in S09.  The Pleiades is omitted because the white dwarf in question is very young, the progenitor was very massive, and the theoretical errors in \mtp\ are large.  We have updated the S09 WD measurements or cluster parameters (such as age or metallicity) for several clusters in our sample; we discuss the adopted WD and cluster data in Section~\ref{sec:mf} and Section~\ref{sec:mi}, respectively.

\subsection{White Dwarf Samples and Final Masses}\label{sec:mf}

Final white dwarf masses are determined from spectra of their Balmer lines, which are particularly sensitive to changes in surface gravities ($g$) and effective temperature (\teff).  Theoretical WD cooling curves provide a mass-radius relationship which is used to infer the final white dwarf mass and the cooling age (\tcool) of the remnant, with only a mild dependence on the WD composition.

S09 use the cooling curves of \citet{Salaris00} to determine \mf\ and \tcool\ and their uncertainties from these data. Investigating the systematic uncertainties inherent in the determination of the IFMR was a major focus of S09. As such, their reported error in \mf\ incorporates not only the uncertainty from observational errors but also the resulting range of \mf\ calculated using alternative cooling tracks from \citet{Althaus03} and limiting case choices of: neutrino energy loss rates, conductive opacities, WD core composition, and hydrogen envelope thickness. The average fractional error in final mass and cooling age for the S09 sample is $8.4\%$ and $34.9\%$, respectively. Note that these errors are extremely conservative and incorporate many more potential error sources than in typical IFMR studies.

We adopt the WD masses and cooling ages from S09 for six of the nine clusters in our sample. For these clusters, S09 has compiled the most recent and high resolution data. S09 adopt the \logg\ and \teff\ measurements first compiled in \citet{Ferrario05} for the Hyades \citep{Claver01}, NGC 2516 \citep{Koester96}, and M37 \citep{Kalirai05}. WD measurements from NGC 6819 and NGC 7789 \citep{Kalirai08}; and NGC1039 \citep{Rubin08} complete the list of clusters whose WD data is up to date in S09.

There are more recent WD observations in NGC 3532, M35, and Praesepe. In NGC3532, \citet{Dobbie09} present high resolution spectroscopic and photometric observations of six WDs, including the three originally observed in \citet{Koester93} and reported in S09. With precise distance modulus measurements, they determine that two of these six WDs are not associated with the cluster. The final masses and cooling times for the remaining four WDs are interpolated from the theoretical cooling curves of \citet{Fontaine01}. The average fractional error in final mass and cooling age is $6.0\%$ and $22.2\%$, respectively.

\citet{Williams09} expand upon their previous study \citep[][reported in S09]{Williams04} of WDs in M35. They present high resolution spectroscopic observations and updated \logg\ and \teff\ measurements of 12 DA WDs. Three WDs in their sample were not fit satisfactorily by model atmospheres (possibly due to magnetic fields) and we remove them from our sample. \citet{Williams09} interpolate the WD cooling curves of \citet{Fontaine01} to obtain masses and cooling times for the WDs in their sample. The mean fractional error in final mass for the nine WDs we analyze is $9.3\%$ and is $51\%$ in cooling time.

\citet{Casewell09} present high-resolution spectroscopic observations of 9 WDs in Praesepe using the Very Large Telescope. They determine that the object WD0836+201 has been mislabeled by previous studies and has strong magnetic fields. Using radial velocity measurements, they argue against the inclusion of another candidate, WD0837+218, as a cluster member. We use their remaining seven candidates as our Praesepe WDs. Taking advantage of their high resolution, \citet{Casewell09} measured line core velocity shifts of the $H\alpha$ and $H\beta$ lines and refit model atmospheres to the spectra accounting for these shifts, resulting in lower $\chi^2$ fits. \citet{Casewell09} interpolate their measured \teff\ and \logg\ amongst a grid of cooling curve models from \citet{Fontaine01} to determine the WD masses and cooling times. The average fractional uncertainty is $5.6\%$ in WD mass and $11.6\%$ in cooling age.

We note our adopted WD masses from these sources are calculated using a different set of theoretical cooling curves from the rest of our sample. However, the uncertainty in WD mass is dominated by observational errors rather than systematic ones stemming from the choice of WD cooling curves \citep{Salaris09}. On the other hand, WD cooling ages vary more when interpolated using different models. To account for this systematic uncertainty, we assume a fractional error of $50\%$ for the WD cooling ages obtained using the \citet{Fontaine01} models. For nominal choices of WD composition, envelope thickness, energy loss rates, and opacities, the fractional uncertainty in WD cooling ages for the remainder of our sample is still $<20\%$ in most cases. Improvements in the observational measurements of \logg\ and \teff\ and the removal of cluster non-members from the sample more than offset any systematic affect stemming from the inclusion of a second group of WD cooling curve models in our analysis.

\subsection{Cluster Parameters}\label{sec:cluspar}

Progenitor lifetimes are a function of WD cooling ages and cluster ages. The accuracy of our initial-final mass data set depends critically on constraining the cluster distance and age. Although a few clusters have measured parallaxes, the distances to most are inferred by main sequence fitting.  Photometric methods can be used to infer composition, reddening, and distance even for clusters with limited membership and spectroscopic data (see for example Pinsonneault et al. 1997, 2004; An et al. 2007).  Well-studied open clusters can have spectroscopic metallicity measurements, extinction inferred from polarization studies, and both radial velocity and proper motion membership data.  The uncertainties in these basic cluster parameters are largely determined by the available information in each specific cluster, which we discuss below, and set the uncertainty in the inferred cluster distance. The measured distance to the cluster yields the turnoff luminosity, which can be combined with isochrones to determine the age and the mass-main sequence lifetime relationship (see S09 for a discussion).  The errors in these two aspects are distinct in nature.

Core size deserves special comment.  Convective core overshooting \citep{Chiosi86} or rotational mixing \citep{Maeder00} are difficult to model theoretically, and both mechanisms have the practical effect of extending the main sequence lifetime by providing extra fuel.  The main sequence is broader than that predicted by models without overshoot or mixing, which is evidence that this phenomenon is real to some degree \citep[\eg][]{Andersen91, Torres10}.  As a result, the main sequence lifetime-mass relationship has substantial theoretical uncertainties.  However, the total fuel burned is more reliable than the cluster ages because the helium core mass at the end of the main sequence is less than the minimum core mass required for helium ignition.  Uncertainties related to overshooting will thus be important for intermediate mass stars, where the star leaves the main sequence with a helium core greater than the minimum required for core He burning, but the impact on stars at the lower end of the range will be reduced.  For the purposes of this paper we adopt a conservative limiting case approach, independently inferring fuel consumption bounds from models with and without overshoot.

\subsection{Initial Masses}\label{sec:mi}

The progenitor lifetime is the difference between the cluster age (\tclus) and the cooling age of the WD (\tcool). Stellar evolutionary tracks spanning a range of stellar mass and metallicity are then interpolated to yield the best-fitting progenitor mass (\mi).  WD spectra modeled by theoretical cooling curves constrain \tcool\ (Section~\ref{sec:mf}). Set by the distance to the cluster and metallicity, the cluster turn-off luminosity is a direct 
indicator of \tclus. The cluster color-magnitude diagram (CMD), $[Fe/H]$, and $E(B-V)$ are the three main data inputs in main sequence fitting algorithms used to constrain the distance modulus to the cluster. S09 first assume, using measurements from the literature, a metallicity and reddening for each cluster (see their Table 2 and 3 for their sources and values, respectively). To obtain the cluster distance, they empirically fit the main sequence of $V$-$(B-V)$ CMDs using a sample of of field dwarfs with known metallicities and Hipparcos parallaxes \citep{Percival03}. Once the distance is known, the turn-off luminosity is measured. S09 interpolate the isochrones (both with and without convective overshooting) of \citet{Pietrinferni04} in both turn-off luminosity and metallicity to constrain the best fit cluster age.

We adopt the cluster parameters, namely cluster age and composition, presented in S09 for three of the nine clusters in our analysis. The composition measurements included in S09 of the Hyades, NGC 6819, and NGC 1039 are either taken from the most recent, high resolution studies of these clusters or are in agreement with current measurements. Additionally, the computed distances to these clusters align with previous distance investigations (\eg\ \citealt{Perryman98}; \citealt{Kalirai01}; and \citealt{Jones96} for the Hyades, NGC 6819, and NGC 1039 respectively).

We searched the literature for revised metallicity measurements of and/or distance determinations to all the clusters in our sample. For six of the clusters, we modify the cluster composition, distance, or both from the values found in S09. In two of these six clusters, S09 either had difficulty with their main sequence (MS) fitting technique due to poor data (in the case of NGC 3532) or chose a reddening value significantly different than that calculated in the recent literature (M37). In other clusters in our sample, several high resolution spectroscopic studies over the last decade have produced very precise metallicity measurements. We use these more recent values in our final calculations. 

We put all of our cluster ages on the same relative scale. \citet{Percival03} present the main sequence fitting technique applied in S09, and derive a metallicity dependency of $\Delta (B-V)= 0.154\Delta[Fe/H]$ for their procedure. Their algorithm produces distances relative to an assumed distance to the Hyades, principally by matching the shape of the Hyades MS to the cluster CMDs. Assuming a linear slope with a magnitude of $\sim5$ for the relevant portion of the Hyades MS \citep{Percival03}, $\Delta (B-V)= \Delta M_v/5$; therefore, $\Delta M_v = 0.77*\Delta[Fe/H]$. \citet{Twarog09} performed an investigation of MS fitting techniques using nearby field stars with Hipparcos parallaxes and precise metallicities and find  $\Delta M_v = 0.98*\Delta[Fe/H]$. The metallicity dependency determined by \citet{Percival03} and  \citet{Twarog09} are similar; implementing the \citet{Twarog09} relation would not significantly impact our results. We modify cluster ages corresponding to changes in distance according to the errors presented in S09. We assume that, for a given cluster, the fractional error in distance is equal to the fractional error in age. If $\Delta(m-M)$ is the S09 cluster distance subtracted from the new distance then $\Delta (m-M)/\sigma_{(m-M)}* \sigma_{t_{clus}}=\Delta t_{clus}$, where $\sigma_{(m-M)}$ and $\sigma_{\tclus}$ are from S09 and the new cluster age is $t_{clus}=t_{clus, S09} - \Delta t_{clus}$. We adopt ages relative to those of S09 for consistency. Below we summarize our revisions to these six clusters.

\textbf{Praesepe} \citet{An08} present a new distance measurement to this cluster using empirically calibrated isochrones. Additionally, they obtain high signal to noise spectroscopy of several Praesepe stars, reporting \feh$=0.11\pm0.03$. This is between the \feh$=0.04\pm0.06$ from the literature compilation of \citet{Gratton00} used by S09 and the subsequent higher \feh$=0.27\pm0.10$ found by \citet{Pace08}. \citet{An08} find $(m-M)_0=6.33\pm0.04$ assuming \feh$=0.14\pm0.02$ (a weighted mean of their result and literature values excluding non-members) and $E(B-V)=0.006\pm0.002$. We adopt the latter \citet{An07} value for our metallicity. Comparing the metallicity used in this distance determination with that of S09, we find $\Delta\feh=0.14-0.04=0.10$. Accounting for this change in metallicity according to the metallicity dependence in \citet{Percival03}, the de-reddened S09 distance is $(m-M)_0=6.32 \pm 0.04$. The \citet{An08} distance is $0.01\ mag$ larger than expected, implying a relatively younger cluster age (\tclus) of $637\pm50\ Myr$ using isochrones incorporating convective overshoot (OS) and $440\pm40\ Myr$ assuming no overshoot (nOS). The errors in the cluster age remain unchanged.

\textbf{NGC 2516} \citet{An08} simultaneously best-fit their photometry of NGC 2516 with \feh$=-0.04\pm0.05$, $E(B-V)=0.117\pm0.002$, and $(m-M)_0=8.03\pm0.04$. This represents a substantial improvement in $\sigma_{\feh}$ over S09 ($\sigma_{\feh}=0.11$). De-reddening and accounting for $\Delta\feh=0.12$, the S09 distance is \dm$_{0}=8.04$. Hence, $\Delta\dm_0=-0.01\ mag$ yielding lifetimes of $\tau_{OS}=137\pm29\ Myr$ and $\tau_{NOS}=91\pm26\ Myr$. The error in cluster age has been reduced from their S09 values by the factor $\sigma_{\dm,An08}/\sigma_{\dm, S09}=0.57$.

\textbf{NGC 3532} S09 could not find sufficiently precise photometry of this cluster in the magnitude range necessary to match their WD templates; subsequently, they had to scale the derived distance and age from their results for Praesepe. Assuming the same reddening as S09, \citet{Kharchenko05} find $(m-M)_V=8.61\pm0.2$ using optical photometry and Hipparcos proper motions. Their distance measurement is $0.21\ mag$ larger than S09, giving $\tau_{OS}=316\pm80\ Myr$ and $\tau_{NOS}=216 \pm 80\ Myr$. Note that the errors in cluster age are reduced by $\sigma_{\dm,K05}/\sigma_{\dm, S09}=0.80$. We could not find a a recent, high resolution spectroscopic metallicity determination of this cluster. We use the S09 value: \feh$=0.02\pm0.06$ \citep{Gratton00}.

\textbf{M37} S09 fit the main sequence of this cluster using two different sets of cluster parameters. The first, with super-solar metallicity and $E(B-V)=0.30$ produced a cluster age that was improbably young. The progenitor masses for this age were so large relative to the remnant masses that the theoretically predicted core mass at the onset of the TP-AGB phase was typically greater than the remnant mass. S09 tried a substantially lower metallicity, \feh$=-0.20$, and reddening, $E(B-V)=0.23$ and obtained a more reasonable cluster age in agreement with previous age determinations. We scale our results with this latter cluster characteristic set. \citet{Hartman08} performed high resolution spectroscopic and photometric observations of M37 using the MMT. They determine \feh=$0.045\pm0.044$, $E(B-V)=0.227\pm0.038$, and \dm$_V=11.57\pm0.13$. Comparing this distance with that of S09, $\Delta\dm_0=0.17$. However, adjusting the S09 distance to the same metallicity ($\Delta\feh=0.24$), we find \dm$_V=11.58$. Thus, $\Delta\dm_0=11.57-11.58=-0.01\ mag$, and the cluster age for M37 is $\tau_{OS}=554\pm54\ Myr$ and $\tau_{NOS}=354\pm43\ Myr$. The errors in cluster age have increased by the factor $\sigma_{\dm,An08}/\sigma_{\dm, S09}=1.083$ over their S09 counterparts.

\textbf{M35} We found one high resolution, high signal to noise measurement of M35's composition (\feh$=-0.21\pm0.10$) by \citet{Barrado01}. In M35, $\Delta\feh=-0.21 - (-0.19)=-0.02$ and thus $\Delta\dm=-0.015$. The revised cluster ages are slightly older than those in S09: $\tau_{OS}=124\pm30\ Myr$ and $\tau_{NOS}=88\pm 25\ Myr$. Cluster age errors remain unchanged from the S09 values.

\textbf{NGC 7789} \citet{Tautvaisiene05} determine \feh$=-0.04 \pm 0.05$ in NGC 7789 using high resolution spectra of evolved cluster members while the measurement (\feh$=-0.13 \pm 0.08$) quoted in SO9 is an average of photometric and low dispersion spectroscopic abundance measurements \citep{Gratton00}. $\Delta\feh=-0.04-(-0.13)=0.09$; the accompanying change in distance is $\Delta\dm=0.07$. The cluster age is $58\ Myr$ younger than stated in S09: $\tau_{OS}=1442\pm100\ Myr$ and $\tau_{NOS}=1042\pm100\ Myr$. Cluster age errors are repeated from S09.

We note that the largest departure in cluster age from S09 is $21\%$. This maximum deviation occurs in the case of NGC 3532, which has more limited data in S09 than any other cluster in the sample. Our adopted age for NGC 3532 is consistent with that of \citet[][$\tau_{OS}=302\pm154\ Myr$]{Koester93}. Typically, however, our cluster ages represent modifications to those of S09 on the order of a few percent. To reduce our error budget, we used more recent cluster composition determinations for many of these clusters. We have demonstrated that our choices of cluster metallicity do not seriously impact our derived cluster ages. Our results would not have changed signifigantly had we simply adopted all the reported cluster ages in S09. 

For each WD, the lifetime of its progenitor is $\tprog=\tclus - \tcool$. Errors are propagated from \tcool\ and \tclus\ to \tprog. Using the stellar evolutionary models of \citet{Pietrinferni04}, we construct a data cube of \mi\ as a function of \tprog\ and \feh. For the six clusters with modified cluster ages or WD cooling ages, we interpolate the cube linearly in \feh\ and quadratically in \tprog\ to obtain the progenitor mass. For each object, we interpolate this grid $2\times10^6$ times, each time independently sampling a Gaussian distribution in \mi\ and \feh\ (the width of each Gaussian is $\sigma_{\tprog}$ and $\sigma_{\feh}$, respectively). One sigma error bars encapsulate the central $68\%$ of the values interpolated from the grid. The uncertainty in \mi\ incorporates the error in \tprog\ and \feh. S09 interpolate the \citet{Pietrinferni04} models in the same fashion to obtain progenitor masses for WDs in the other three clusters. In their error budget, they also include the systematic offset in \mi\ from interpolating a different stellar evolutionary model set (namely those of the PADOVA group \citep{Girardi00}). While our revised \mi\ measurements in Praesepe, NGC 2516, NGC 3532, M37, M35, and NGC 7789  do not include the systematic error associated with using the PADOVA models, this potential contribution to the total $\sigma_{\mi}$ is small when compared to the cluster parameter uncertainties (S09).

\subsection{Cluster Averages}\label{sec:averages}

The final and initial masses of all 48 WDs in our sample are shown in Figure~\ref{fig:ifmr}. In many of the clusters, there is considerable scatter in these values, especially when \mi$> 3 M_{\odot}$. When $\tclus - \tcool$ is small, errors in either quantity contribute more strongly to \tprog, resulting in a spread of initial masses for a single cluster. We reduce this scatter by representing each cluster by its weighted mean \mi\ and \mf\ \citep[first proposed in][]{Kalirai08}. To determine the influence of outlying measurements, we also compute the median \mi\ and \mf\ for each cluster and adopt the standard error approximation $\sigma_{med}=\pi/2*\sigma_{mean}$.  We find no statistically significant change in our results when adopting the median \mi, \mf\ for each cluster instead of the mean. For the remainder of this paper we present our results using the weighted mean \mi\ and \mf\ of each cluster.  

\subsection{Fraction of Core Growth in the TP-AGB Phase}\label{sec:m1tp}

The core mass at the first thermal pulse of the AGB stage is fixed by theory via the star's initial mass and composition. Similar to the calculation of \mi\ from \tprog, we construct a data cube of the core mass at the first thermal pulse (\mtp) given \mi\ and \feh\ using the \citet{Pietrinferni04} models. \mtp\ is interpolated linearly in \feh\ and quadratically in \mi. Central values and errors are calculated similarly to that of \mi\ in Section~\ref{sec:mi}. The error in \mtp\ is the one sigma range of interpolated \mtp\ given the uncertainties in cluster \feh\ and \mi. As the final remnant mass is equal to the core mass at the tip of the AGB, the core grows by \delmc$=\mf-\mtp$ in the TP-AGB phase. The uncertainties in \mtp\ and \mf\ are propagated to determine $\sigma_{\Delta M_c}$.  The fractional contribution of the TP-AGB phase to final core mass (\fctp) is then $\delmc/\mf$. Table~\ref{tab:fuel} lists the \mi, \mf, \delmc, and \fctp\ of each cluster. 

\subsection{Fuel Consumption: Mass-Light Coupling}\label{sec:fuel}

The total energy during any evolutionary phase of a star's life is directly proportional to the fuel consumed \citep{Renzini86}. As nuclear burning is a source of energy for stars, the core's growth during the TP-AGB phase represents a direct lower bound on the fuel burned during this phase \citep{Marigo01}. True, there are contributions to the light output from gravitational contraction of the core and neutrino loses. However, these can be ignored in our analysis as they only represent corrections on the percent level \citep{Marigo01}. In addition, significant observable light will come from nuclear burning reactions whose products (notably He) are expelled in stellar winds and do not end up in the remnant. Hence, we define the growth of the core as a strict \emph{lower bound} on the fuel consumed in the TP-AGB phase.

Mathematically, we couple the consumed fuel to the energy released during the TP-AGB phase via \citep[\cf\ eq. 5 from][]{Marigo01}: 
\begin{equation}\label{eq:fuel}
F(M_\mathrm{f}, M_\mathrm{i})=\frac{1}{A_H}\int L_{M_i}(t)dt.
\end{equation}
The fuel, $F(M_i)$, is expressed in solar masses and the integral is the sum of the energy released during the TP-AGB phase. The conversion from energy to mass is represented by $A_H$, the efficiency of $H$ burning reactions. Here, we adopt $A_H= 9.75 \times 10^{10}\ L_{\odot}\ yr\ M_{\odot}^{-1}$ following \citet{Marigo01}. Above, we note that nucleosynthesis is the dominant stellar energy source. The fuel consumed in the TP-AGB phase is therefore:
\begin{equation}\label{eq:fueltp}
F_{TP-AGB}\simeq (X_{1,2} +0.1)(\Delta M_c) + M_y^{TP-AGB}(He) + (1.1-Y')M_y^P(CO).
\end{equation}
$X_{1,2}$ is the surface mass fraction of hydrogen after the second dredge-up event, $Y'$ is defined as $Y_{1,2}/(X_{1,2}+ Y_{1,2})$ where $Y_{1,2}$ is the mass fraction of helium in the envelope after the second dredge-up, and $M_y^{TP-AGB}(He)$ and $M_y^P(CO)$ refer to the stellar yield of $He$ and $CO$, respectively, during the TP-AGB phase. We assume an initial H abundance ($X_0$) of $0.71$ and map to $X_{1,2}$ using the tabulated surface compositions in \citet{Bertelli08, Bertelli09}. Irrespective of our lack of knowledge regarding the yields of He and CO we can construct our strict lower limit using only the observed growth of the core during the TP-AGB phase. The \emph{minimum} bound on the fuel consumed during this phase, and hence the energy released, is obtained by setting the fuel in equation~\ref{eq:fueltp} equal to the first term on the right hand side and setting that equal to $F(\mi)$ in equation~\ref{eq:fuel}. Solving for $\int L_{M_i}(t)dt$ in equation~\ref{eq:fuel}, we obtain the minimum energy release (\lmin) necessary to increase the core mass by \delmc. The uncertainty in  \lmin\ is propagated from $\sigma_{F_{TP-AGB}}$. We list this value for each cluster in Table~\ref{tab:fuel}.

We compare \lmin\ to the predicted integrated luminosity of evolutionary tracks incorporating the TP-AGB phase \citep[\eg][]{Pietrinferni04, Marigo07, Bertelli08, Bertelli09}. Straightforwardly, we can integrate $\int L_{M_i}(t)dt$ in the TP-AGB phase for any choice of model and find the predicted energy release. As they incorporate up to date physics and produce tracks that span a wide range of metallicity, we use the latest TP-AGB models from the PADOVA group in our analysis \citep{Bertelli08, Bertelli09}. The range of composition in our cluster sample is covered by their scaled-solar tracks with (Z=0.008, Y=0.26), (Z=0.017, Y=0.26), and (Z=0.40, Y=0.30). We create a grid of $\int L(t)dt$  in the TP-AGB phase as a function of \mi\ and composition. We interpolate the grid quadratically in mass and lineally in metallicity to determine the total predicted luminosity output (\lmod).  The procedure to calculate the central value for \lmod\ and $\sigma_{\lmod}$ is similar to the calculation of \mi\ and its error (Section~\ref{sec:mi}). The ratio $f(\lmod)=\lmin / \lmod$ represents the fraction of predicted energy released in the TP-AGB phase accounted for by core growth. The quantity $1- f(\lmod)$ is now the fraction of the total predicted energy that is traced by the products of relic Helium and Carbon burning released to the ISM during the TP-AGB phase. Given either theoretical or observational constraints on the total light output in the TP-AGB phase, our analysis leads to direct predictions of TP-AGB stellar yields. 

In the above prescription we neglect neutrino loses and the the contribution of the core's gravitational contraction to the energy budget of the TP-AGB phase. Gravitational contraction can provide, at most, $\sim5\%$ of the energy budget of the TP-AGB phase. Nucleosynthesis of H into He and eventually C and O has an efficiency of $\sim5\times10^{18} erg/g$ whereas the gravitational energy released as luminous radiation per unit mass from dwarf creation is $\sim2\times10^{17} erg/g$. Energy generation from core contraction is negligible given the uncertainties in the dominant energy source- nucleosynthesis. We determine the energy output accounted for by the growth of the core and establish constraints on the stellar yields of He, C, and O during the TP-AGB phase as a function of initial stellar mass in Sections~\ref{sec:corefuel} and ~\ref{sec:Hefuel}, respectively.

\section{Results}\label{sec:results}
Observations of white dwarfs in open clusters combined with established evolution models constrain the fuel consumption, and hence light output, during the TP-AGB phase. Our results demonstrate that the stellar core grows by a non-negligible amount during the TP-AGB phase, regardless of progenitor mass. Below, we address the evolution of core mass growth in TP-AGB stars as a function of initial mass, couple this core mass increase to a lower bound on TP-AGB star light output, compare our results with recent theoretical models of the TP-AGB phase, and discuss how our results change with different overshooting conditions.

\subsection{Initial Final Mass Relation}\label{sec:ifmr}
\begin{figure}
\plotone{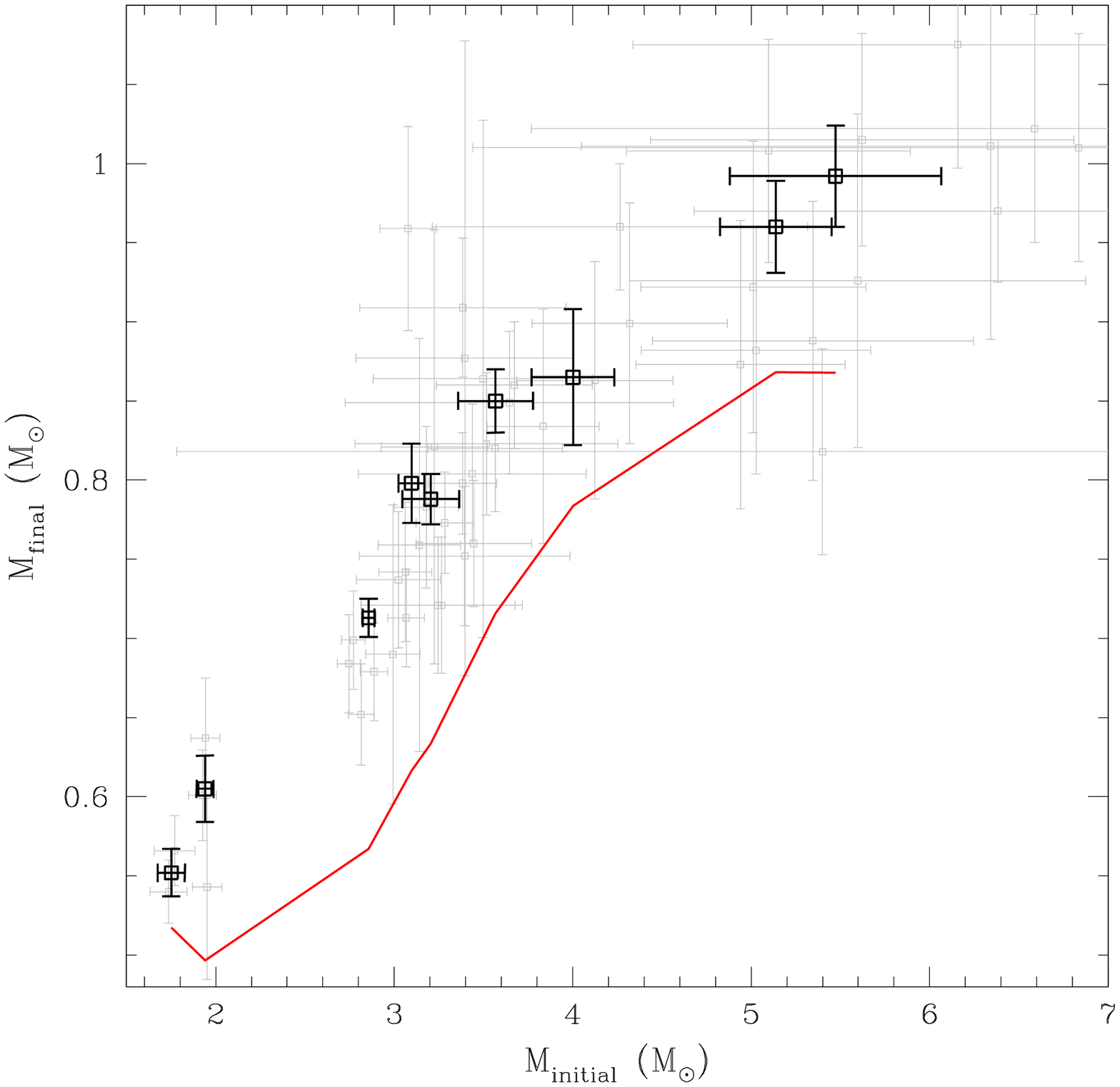}
\caption{\label{fig:ifmr} The IFMR for the 48 WDs in our sample. The progenitor masses are calculated using cluster ages and stellar evolutionary theory that incorporate a degree of convective overshooting \citep{Pietrinferni04}. The \mf-\mi\ pairs for each WD are in light gray; error bars represent one $\sigma$ deviations in each quantity. The open black squares show the weighted mean \mf\ and \mi\ for the nine clusters in our sample. Error bars are the one $\sigma$ error of the means. Given an initial mass and cluster metallicity, theory predicts the mass of the core at the first thermal pulse (\mtp). The solid red line connects the nine \mtp\ - \mi\ points created using the weighted mean \mi\ and the \feh\ of each cluster. The observed remnant masses in these clusters all lie above the red line. As \mf\ is the mass of the core at the tip of the AGB, stellar cores must grow during the TP-AGB phase.
}
\end{figure}

We plot the IFMR for our sample in Figure~\ref{fig:ifmr}. The observed, final masses and their calculated progenitor masses are represented by the gray points. The weighted mean \mf-\mi\ pair for each cluster is in black. When compared with the literature from which the initial data came, our final masses are within the errors of earlier measurements. Typically, the final masses of a given cluster are consistent across different IFMR studies despite coming from various theory groups \citep[\eg][]{Williams09}. On the other hand, the cluster parameters that dictate the calculated initial masses are less well constrained;  cluster age, in particular, is notoriously difficult to determine. The uncertainties in the inputs to MS isochrone fitting: namely composition, reddening, and distance, all contribute to uncertainty in the cluster age and hence, progenitor masses. Progenitor masses calculated in our analysis rather than in S09 still show good agreement with those of other studies assuming the same cluster ages, \eg, in M37 our weighted mean \mi$=3.21\pm0.11 \msol$ for \tclus$=563\pm31\ Myr$ and \citet{Ferrario05}, who use \tclus$=520\pm80\ Myr$,  find \mi$=3.27\pm0.12 \msol$. Irrespective of the assumption of convective overshooting, the average fractional uncertainty in our sample for \mf(\mi) is $3\% (7\%)$. 

The red line in Figure~\ref{fig:ifmr} connects the anticipated core mass at the first thermal pulse (\mtp) for each cluster's weighted mean initial mass. We determine \mtp\ given \mi\ and \feh\ according to \citet{Pietrinferni04} (see Section~\ref{sec:m1tp}). The core mass at the first thermal pulse is a monotonically increasing function of initial mass for a given metallicity. The clusters in our sample span $0.32$ dex in \feh. The differing compositions of the two clusters with the lowest progenitor mass explains the small dip in the red line at \mi$\sim 2 M_{\odot}$. The relationship between the core mass at the first thermal pulse and the progenitor mass is a robust theoretical prediction; comparison between the models used here and those of \citet{Girardi00} yield identical results to within $2\%$. In Figure~\ref{fig:ifmr}, all the cluster mean final masses are larger than the predicted core mass at the first thermal pulse. Given nominal assumptions regarding stellar evolution, current cluster observations demand that the core grows substantially during the TP-AGB phase; thus, TP-AGB stars must contribute significantly to cluster luminosity. While this idea has been gaining traction in the literature \citep[\eg][]{Maraston06}, we show, using the IFMR and accepted stellar interior theory, that TP-AGB stars must be considered a large luminosity source in population synthesis models.

\subsection{Core Growth in the TP-AGB Phase}\label{sec:frac}
\begin{figure}
\plotone{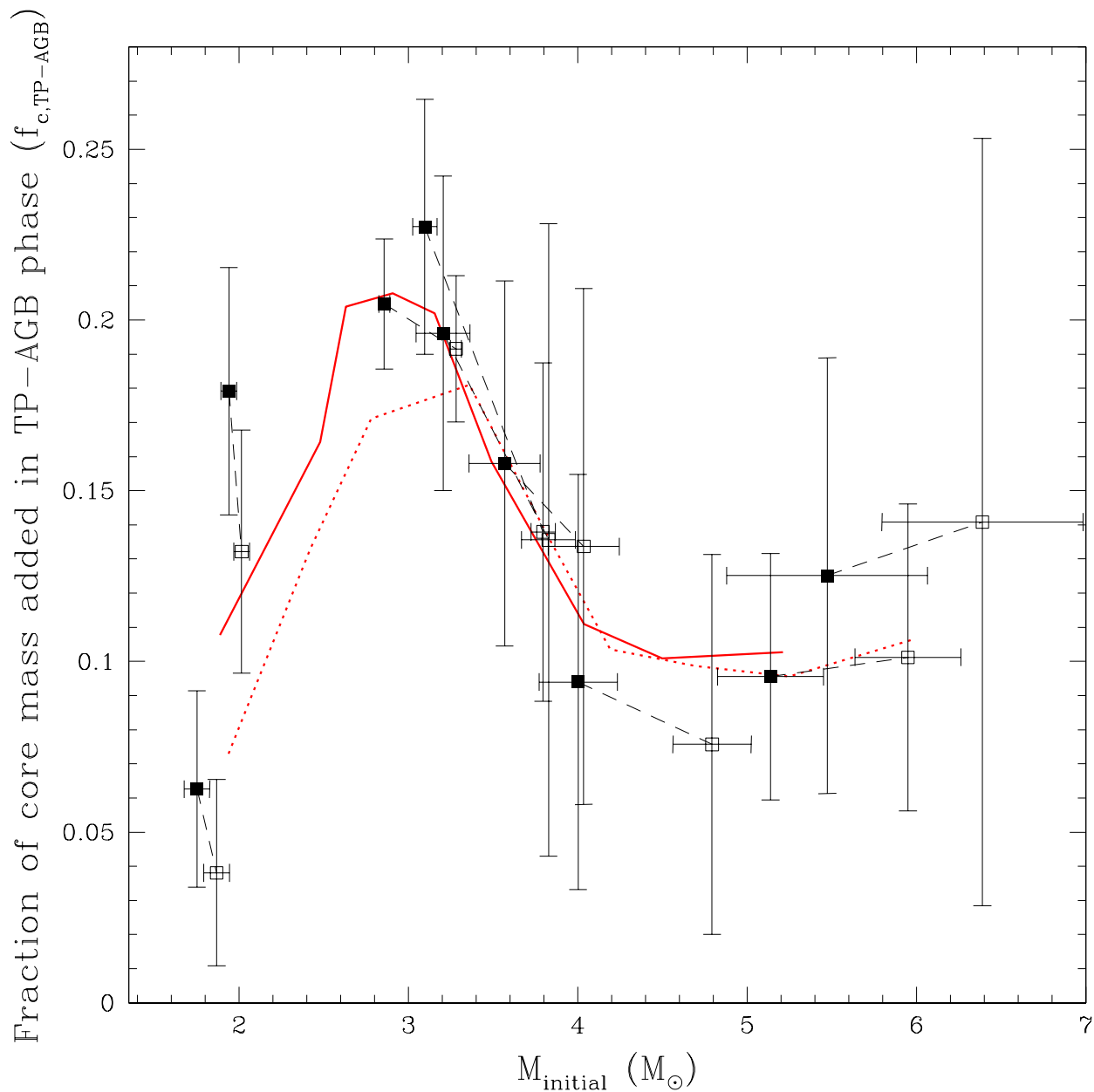}
\caption{\label{fig:frac} The fraction of the final remnant mass built up in the TP-AGB phase as a function of initial mass. The squares are calculated using the weighted mean initial and final masses of each cluster. Filled squares use evolution theory and isochrones incorporating convective overshooting; open squares represent the limiting case of no overshooting. The dashed lines connect these two choices for each cluster. The  solid (dashed) red line connects the results of a moving weighted mean of the filled (open) squares. The fractional core mass growth in the TP-AGB phase has a broad peak between $2$-$3\msol$. }
\end{figure}

The core growth during the TP-AGB phase is a strict lower limit on the mass of nuclear burning products in TP-AGB stars and the requisite contribution to stellar populations' bolometric luminosity. In Figure~\ref{fig:frac}, we illustrate the fraction of the final core mass grown during the TP-AGB phase as a function of initial mass. We define this fractional mass as \fctp$=\delmc / \mf$ where \delmc\ is the difference between the final remnant mass and the core mass at the onset of the TP-AGB phase. The data points are made using the weighted mean initial and final masses for each cluster. As mentioned in Section~\ref{sec:methods} we  compute the progenitor mass and \mtp\ using evolutionary tracks both with and without overshooting. Filled squares represent masses found with models and cluster ages incorporating overshoot; open squares are masses that do not take overshooting into account. Dashed lines connect the results of these two cases for each cluster. The error bars, calculated according to Sections~\ref{sec:m1tp} and ~\ref{sec:mi}, represent one sigma errors in fractional core mass gained during the TP-AGB phase and initial mass, respectively. The red lines connect the moving weighted mean of each data point type, illustrating the overall trends in the figure. At every filled or open point, we calculate the weighted mean of itself and its closest neighbor in both directions of \mi\ (end points are only averaged with the closest data point). The solid red line connects the averages of the filled data points (OS); the dashed red line connects the open square (nOS) averages.

These moving averages highlight the large increase in fractional core mass growth during the TP-AGB phase between $2$ and $\sim3 M_{\odot}$ (up to $4 M_{\odot}$ in the no overshooting case). In the case where convective overshooting is considered (OS), the weighted mean \fctp$= 0.20 \pm 0.01$ when $1.9 \leq\mi\leq 3.6$ and drops by a factor of two to $0.08\pm 0.02$ elsewhere. This broad peak in \fctp\ above the remainder of the sample is significant at the  $3.5 \sigma$ level. For progenitor masses calculated with models and cluster ages that do not include convective overshooting (nOS), \fctp\ is generally smaller than its OS counterpart and the fraction of core mass gained in the TP-AGB phase peaks at a slightly higher progenitor mass ($3.5 M_{\odot}$ versus $3 M_{\odot}$). These shifts are a natural consequence of the inclusion or absence of convective overshooting in evolutionary models and isochrones. Convective overshooting models predict a larger He core and longer lifetime along the main sequence compared to nOS models. Due to these larger cores, \mtp\ is larger for a given \mi\ when overshooting is considered. If the initial masses calculated with and without convective overshooting were the same, \fctp\ would be larger in the nOS case. However, OS isochrones assign older cluster ages (and subsequently smaller \mi) than nOS isochrones given the same cluster CMD. The \mtp\ - \mi\ relation is monotonically increasing and thus the nOS \mtp\ is higher than its OS counterpart in the same cluster when \mi$\leq 5 \msol$ (the \mtp\ - \mi\ relation flattens slightly at higher \mi; nOS \mtp\ are similar to the OS \mtp\ in this mass range). Still, the nOS average curve shows the same shape as that composed of OS progenitor masses. On a cluster by cluster basis, the OS and nOS core growth fractions are typically within $5\%$ of each other. While observations of gaps in open cluster CMDs and the HR diagram postions of binary stars give more credance to models invoking core overshoot, the precise extent of convective cores is still a matter of active debate. It is thus encouraging that our results are surprisingly independent of this signifigant theoretical uncertainty. Given the similarity of our results regardless of overshooting behavior and the facts that models with convective overshooting are favored, future sections will only address our OS results.

Interestingly, the rise in \fctp\ roughly coincides with the peak in the lifetime function predicted by recent TP-AGB models \citep[\eg][]{Marigo07}. The lifetime of the TP-AGB phase is set by the time between the first thermal pulse and the removal of the envelope. The envelope can be removed via mass loss to the ISM or through outward advancement of the H-burning shell (hot bottom burning). In many TP-AGB models, the lifetime of the phase depends critically on the occurrence of the third dredge-up (3DU) episode. In the 3DU, carbon is mixed into the envelope, causing the $C/O$ ratio to rise and the opacity of the envelope to increase increase. In turn this creates and sustains a dust-driven wind that greatly enhances the mass loss rate of the envelope. In \citet{Marigo07}, the 3DU occurs when the mass of the core exceeds a threshold ($M_{\mathrm{crit}}$) that, at solar metallicity, is relatively flat when \mi$<3 \msol$ and becomes a steep function of \mi\ at higher masses. Though \mtp\ typically grows with \mi, the \mtp\ -\mi\ relationship plateaus between $1.5 \msol$ and $3 \msol$. The core must grow fractionally more in the TP-AGB phase to reach $M_{\mathrm{crit}}$ in this mass regime; therefore, TP-AGB star lifetimes are longest for these initial masses. More massive stars (\ie\ $\geq 3 M_{\odot}$) experience their third dredge-up more quickly as their core mass at the onset of the TP-AGB phase is closer to $M_{\mathrm{crit}}$. Conversely, we observe that when \mi$<\sim 1.8 \msol$, stellar cores at the first thermal pulse are already a substantial fraction of $M_{\mathrm{crit}}$. The lowest \mi\ in our sample, $1.75 \msol$, has the smallest \fctp\ and is predicted to have a shorter lifetime than $2\ \msol$ and $3\ \msol$ TP-AGB stars.

\subsection{Fuel from Stellar Core Growth}\label{sec:corefuel}

Using the procedure outlined in Section~\ref{sec:fuel}, we determine the fuel consumption during the TP-AGB phase required by core mass growth. We have already determined the mass added to the core during the TP-AGB phase ($\Delta M_c$) for each cluster. Using equations ~\ref{eq:fueltp} and ~\ref{eq:fuel}, $\Delta M_c$ is converted to its equivalent energy output (\lmin). We have shown that the theoretical uncertainty, dominated by the chosen efficiency of convective overshoot, associated with the core mass at the first thermal pulse is relatively small. Thus, \lmin\ is a strong lower bound on the total energy output of TP-AGB stars stemming almost entirely from observational constraints. We compute \lmin\ as a function of initial mass for each cluster (Table ~\ref{tab:fuel}). Core growth in TP-AGB stars demonstrates that the TP-AGB phase contributes a significant portion of a intermediate mass star's overall light generation. Our results emphasize the strong link between the observables of the IFMR and the lower limit on the integrated light during the lifetime of TP-AGB stars.

We now quantify the level of agreement between \lmin\ and the predicted total light output by current evolutionary models of TP-AGB star. We define \fecore$=\lmin / \lmod$ where $\lmod$ is the predicted total energy in the TP-AGB phase and is obtained by integrating TP-AGB stellar tracks. We choose the latest PADOVA group TP-AGB tracks, produced by \citet{Bertelli08, Bertelli09}, as they make predictions for a wide range of metallicities and incorporate up to date physics. In Table~\ref{tab:fuel}, we list \fecore\ and its error on a cluster by cluster basis. The weighted mean of \fecore\ in our sample is $0.56\pm0.04$. The individual \fecore\ for each cluster is within $2\sigma$ of this mean while all but the cluster with the smallest progenitor mass, NGC 6819, and Praesepe are within $1\sigma$ of this mean. Unless \citet{Bertelli08,Bertelli09} substantially under-represent the energy output of the TP-AGB phase, the IFMR provides an estimate of the total light emitted in the TP-AGB phase accurate to within a factor of $\sim2$. 

\subsection{Fuel from Stellar Yields}\label{sec:Hefuel}

If we make the reasonable assumption (Section~\ref{sec:fuel}) that the available fuel in the star during the TP-AGB phase comes primarily from nuclear burning, the quantity $1- \fecore$ is the fraction of total predicted energy accounted for by nucleosynthetic products that were expelled into the ISM rather than added to the core. Assuming \citet{Bertelli08, Bertelli09} correctly predict the total energy output in TP-AGB stars, the product $(1- \fecore)* \lmod$ is the amount of light unaccounted for by core mass growth. Setting $(1- \fecore)* \lmod =  \int L_{M_i}(t)dt$ in equation~\ref{eq:fuel}, we obtain the available fuel, in solar masses (\fhe), that comes from the stellar yield of the star. In equation~\ref{eq:fueltp}, we make a distinction between the yield of He and that of CO. In practice, the binding energy of He is $1/10$ that of CO; thus, me make the approximation that He is the sole component of the stellar yield. The calculated \fhe\ from each cluster is shown in Table~\ref{tab:fuel}. We find our sample's weighted mean \fhe$=0.08\pm0.01\ \msol$. Every individual measurement of the stellar yield is within two $\sigma$ of this mean. These results indicate that every star born at $\sim1.5\msol < \mi < 6\msol$ deposits $\sim 0.1\ \msol$ of He into the ISM during the TP-AGB phase. We discuss potential ramifications and tests of this prognosis in the Section~\ref{sec:disc}.

\section{Summary and Discussion}\label{sec:disc}

Using the WD IFMR and robust stellar evolutionary theory, we have shown that a significant mass fraction of the final, stellar core is generated during the TP-AGB phase. Through simple fuel consumption arguments, the TP-AGB phase must therefore emit a substantial amount of light- proportional to the aforementioned core growth.  There is measurable core growth during the TP-AGB phase in all nine clusters in our sample. The representative progenitor masses range from $\sim 1.7 < \msol\ < 6$ and they span $\sim 0.30$ dex in \feh. The least constrained theoretical assumption in stellar evolutionary models up to the onset of the TP-AGB phase is the depth of convective overshooting and whether or not it takes place. Our main result is valid over a broad swath of theoretical input parameter space, including the uncertain nature of core overshooting. Therefore, we have constructed a strict lower bound on the light output during the TP-AGB phase as a function of initial mass that is predominantly dependent on observational constraints from white dwarfs and their host cluster properties.

The fractional contribution of TP-AGB stars to the total light emitted over their lifetime peaks when $\mi\ \sim 2-3.5 \msol$. In general, $\sim20\%$ of the core is built during the TP-AGB phase in this mass range, double that of lower and higher mass stars. The rise and fall of the TP-AGB's importance with initial mass generally agrees with the relative lifetimes of TP-AGB stars predicted by current models \citep[\eg][]{Girardi07, Marigo07, Bertelli08, Bertelli09}. The maximum of the TP-AGB fractional energy output at $\sim 2-3.5 \msol$ has strong implications for interpreting the integrated light of elliptical galaxies. The Infrared light from TP-AGB stars will be a one of stronger principle components of the spectral energy distributions of galaxies harboring stellar populations $2$ to $3\ Gyr$ old.

Constraining the importance of the TP-AGB phase via the IFMR becomes increasingly difficult for progenitor masses above $\sim5 \msol$. High mass stars have short lifetimes, implying that $\tclus\ - \tcool$ is a small number. In that case, nominal errors in \tclus\ and \tcool\ have a fractionally larger impact on the uncertainty in progenitor age at high mass than at low mass. Additionally, different convective overshooting efficiencies produce more discrepant core masses at various stages of evolution for higher initial masses. Still, we find compelling evidence that $\sim10\%$ of the final core is built up during the TP-AGB stage at high mass regardless of convective overshoot efficiency. The upper limit in progenitor mass for stars experiencing the TP-AGB phase is still controversial and empirical data is scare \citep{Williams09b}. While this work suggests $5.5\ \msol$ stars experience a TP-AGB phase, the IFMR will provide a strong means to determine the highest initial masses to become TP-AGB stars if progenitor masses in this regime can be better constrained in the future.

Our results have potentially important consequences for chemical evolution models. We determine the total energy output in the TP-AGB phase required by the nuclear reactions that directly lead to core mass growth. Assuming nucleosynthesis is the only source of energy generation in the TP-AGB phase, the only other tracer of energy output would be nucleosynthetic byproducts expelled to the ISM. Any measurement of the total light output of a TP-AGB stellar population, combined with this work, places direct constraints on the yield of that population (principally helium). If the helium yield is significant, it would impact the $\Delta Y/\Delta Z$ relationship. Additional helium in the ISM from TP-AGB stars would alter the luminance and lifetimes of subsequent generations of stars. Assuming the models of \citet{Bertelli08, Bertelli09} correctly predict the total light in the TP-AGB phase, we find that $0.08\pm0.01\ \msol$ of helium are released into the ISM by TP-AGB stars. TP-AGB stars are short lived; as such, observations of TP-AGB populations and limits on their luminosity have proven difficult in galaxies more distant than the Magellanic Clouds. However, the above argument can be reversed: given limiting case assumptions of the helium yield of TP-AGB stars and IFMR data, one can establish both upper and lower bounds on the integrated light from TP-AGB stars.

Early evidence suggests that modern models of TP-AGB stars make realistic predictions for their total light output. Mass addition to the core accounts for $\sim 50\%$ of the integrated luminosity from the TP-AGB phase predicted by \citet{Bertelli08, Bertelli09}. If \fecore\ had exceeded $100\%$, we would conclude that the chosen models underestimated the amount of light emitted by TP-AGB stars. Alternatively, if \fecore was small ($<10\%$), we would calculate a relatively large stellar yield. In the future, empirical evidence describing the behavior of the $\Delta Y/ \Delta Z$ relationship or integrated light measurements of TP-AGB populations will constrain TP-AGB stellar yields and impose tighter restrictions on the upper bound of total energy release. Currently, the \citet{Bertelli08, Bertelli09} models predict an amount of luminous energy release by TP-AGB stars that is in agreement with observations. \citet{Bertelli08, Bertelli09} predict that, at its peak, the TP-AGB phase is responsible for $\sim33\%$ of the total light output of stars with $2.2\ \msol\leq\ \mi\ \leq 2.5 \msol$; this fractional contribution falls to $20\%$ when \mi$=3.5\ \msol$ and is $\sim 10\%$ at \mi$=4.0\ \msol$. Though our current results and errors cannot rule it out, it is unlikely that TP-AGB stars are responsible for as much as $40\%$ of cluster light unless the stellar yields of TP-AGB stars are significantly higher than those predicted in Section~\ref{sec:Hefuel}.

This work suggests additional tests of population synthesis models. The predicted light emission from stellar populations in SPS models provides a upper bound on the expected remnant mass as a function of initial mass. In the future, population synthesis codes should quantify their agreement with the IFMR. Various theoretical question marks in SPS modeling can be answered, or at least constrained, by the IFMR. The degree of convective overshooting in stellar evolutionary models has an appreciable impact on implied progenitor masses and the \mtp$-$\mi\ relationship; hence, for a predicted light output in the TP-AGB phase, there will be a statistically significant difference between the remnant mass functions predicted with different overshooting parameters. If we demand that \mtp\ is such that the core grows significantly during the TP-AGB phase, we can place instructive bounds on convective overshoot prescriptions for some clusters. Additionally, the luminosity of TP-AGB populations represent a major uncertainty in SPS models \citep{Conroy09}. Our results showing core mass growth in the TP-AGB phase as a function of initial mass represent a floor to TP-AGB star luminosity and can already exclude some limiting case models. Obviously, models assuming that TP-AGB stars do not contribute to a population's integrated light can no longer be physically motivated. As mentioned above, this work and assumptions as to the chemical yields of TP-AGB stars is sufficient to constrain TP-AGB light output. These new constraints are valuable to future SPS models and will greatly reduce the uncertainties associated with the one of the most theoretically dubious stellar evolutionary phases..

The most prominent sources of error in this analysis are the WD and cluster observations. Measurements of WD surface gravities and temperatures provide constraints on remnant masses and cooling times while cluster observations yield the cluster age and composition. Uncertainties in theory linking these observations and the parameters of interest are typically small compared to the observational errors (S09). By assuming either perfect WD or cluster measurements and repeating our analysis, we find that uncertainties in these two types of observables contribute similarly to the resulting error in core mass growth during the TP-AGB phase. More precise WD or cluster measurements will provide stricter bounds on the energy output of TP-AGB stars. Another potential method to improve the precision of the IFMR (and our results) would be a survey specifically designed to find young white dwarfs in open clusters. In any given cluster, the youngest white dwarfs would correspond to the oldest progenitor lifetimes, reducing the impact of uncertainties in the WD and cluster observations on the implied initial mass and \mtp. More precise measurements of the IFMR would directly lead to more precise calculations in our procedure.

Determining the metallicity dependence of TP-AGB star energy output is an intriguing next step. However, there are not enough open clusters at extreme metallicities to empirically calibrate such a dependence. We would need to increase the number of clusters in our sample, even if the overall spread in metallicity does not change. With many more clusters, one could reproduce our results in bins of metallicity. After empirically characterizing how the energy output of the TP-AGB phase scales with composition (albeit over a limited range), theory may be able to provide a physical model for TP-AGB star luminosity as a function of initial mass and stellar density. If said model could be extrapolated to low metallicity, it would provide a theoretical constraint on the contribution of TP-AGB stars to the spectral energy distributions of high redshift galaxies. At high metallicity, the lowest mass stars may go directly to the white dwarf stage, missing the core He burning stages and beyond \citep{Kilic07}.  This effect is not currently included in population synthesis models and might be important for the low redshift properties of giant elliptical galaxies. Physically motivated priors on TP-AGB star light output would be a substantial advancement of semi-analytic galaxy evolution models as SPS codes still have widely varied characterizations of these stars.

We have shown that the WD IFMR places powerful constraints on the energy released during the TP-AGB stage of stellar evolution. Future observations of TP-AGB populations will constrain the helium yields of these stars- an important factor for chemical evolution models to consider. Alternatively, nominal assumptions regarding the yield of these stars, in conjunction with this work, result in a narrower range of possible TP-AGB population luminosity than previously considered.

\acknowledgements
We would like to thank David Weinberg for his encouragement and thought-provoking discussion.

\begin{deluxetable}{lccccc}
\tablecolumns{6}
\tablecaption{Cluster Metallicity, Distance, and Age}
\tablehead{
\colhead{Cluster} &
\colhead{\feh} &
\colhead{\dm$_0$ (mag)} &
\colhead{OS Age (Myr)} &
\colhead{nOS Age (Myr)} &
\colhead{References} \\
\colhead{(1)} &
\colhead{(2)} &
\colhead{(3)} &
\colhead{(4)} &
\colhead{(5)} &
\colhead{(6)}
}

\startdata
NGC 6819 & $0.09 \pm 0.03$  & $12.15 \pm 0.20$	& $2000 \pm 200$ & $1500 \pm 200$ & 1,2	\\
NGC 7789 & $-0.04 \pm 0.05$ & $12.19 \pm 0.12$	& $1442 \pm 100$ & $1042 \pm 100$ & 3,2	\\
Hyades   & $0.13 \pm 0.06$  & $3.33 \pm 0.01$	& $640 \pm 40$    & $440 \pm 40$   & 4,2	\\
Praesepe & $0.14 \pm 0.02$  & $6.33 \pm 0.04$	& $637 \pm 50$   &   $440 \pm 40$ & 5	\\
M37      & $0.04 \pm 0.04$  & $10.83 \pm 0.13$	& $554 \pm 54$   & $354 \pm 43$ 	& 6	\\
NGC 3532 & $0.02 \pm 0.06$  & $8.48 \pm 0.20$	& $316 \pm 80$ & $216 \pm 80$	& 4,7	\\
NGC 1039 & $0.07 \pm 0.04$  & $8.58 \pm 0.15$	& $250 \pm 25$   & $150 \pm 30$ 	& 8,2	\\
NGC 2516 & $-0.04 \pm 0.05$ & $8.03 \pm 0.04$	& $137 \pm 29$   &  $91 \pm 26$	& 5	\\
M35      & $-0.21 \pm 0.10$ & $10.48 \pm 0.12$	& $124 \pm 30$   & $88 \pm 25$ 	& 9,2	\\
\enddata													        

\tablecomments{\label{tab:clus}Cluster compositions, distances, and ages. (1): Cluster. (2): \feh. (3): Absolute distance modulus. (4): Cluster age using isochrones with overshooting. (5): Cluster age using isochrones without overshooting. (6): Reference for cluster metallicity and distance; if two references are listed, the first applies to metallicity and the second, distance. Note that the distances in column 3 may be modified from their references if different cluster compositions were adopted.}

\tablerefs{
References:
1.~\citet{Bragaglia01};
2.~\citet{Salaris09}; 
3.~\citet{Tautvaisiene05};
4.~\citet{Gratton00}; 
5.~\citet{An07};
6.~\citet{Hartman08};
7.~\citet{Kharchenko05}; 
8.~\citet{Schuler03};
9.~\citet{Barrado01}.
}
\end{deluxetable}

\begin{deluxetable}{lccccccc}
\tablewidth{6 in}
\tabletypesize{\tiny}
\tablecolumns{8}
\tablecaption{Fractional Core Growth, Corresponding Energy Output, and Estimated Helium Yields}
\tablehead{
\colhead{Cluster} &
\colhead{\mi\ (\msol)} &
\colhead{\mf\ (\msol)} &
\colhead{$f_{c,TP-AGB}$ } &
\colhead{\delmc\ (\msol)} &
\colhead{\lmin\ ($10^9 L_{\odot}yr$)} &
\colhead{\fecore} &
\colhead{\fhe\ (\msol)} \\
\colhead{(1)} &
\colhead{(2)} &
\colhead{(3)} &
\colhead{(4)} &
\colhead{(5)} &
\colhead{(6)} &
\colhead{(7)} &
\colhead{(8)} 
}

\startdata
NGC 6819 & $1.75 \pm 0.08$ & $0.55 \pm  0.02$ & $0.06 \pm 0.03$ & $0.03 \pm 0.02$ & $2.66 \pm 1.22$ & $0.29 \pm 0.14$ & $0.07 \pm 0.01$  \\
NGC 7789 & $1.94 \pm 0.05$ & $0.61 \pm  0.02$ & $0.18 \pm 0.04$ & $0.11 \pm 0.02$ & $8.29 \pm 1.65$ & $0.56 \pm 0.14$ & $0.07 \pm 0.02$  \\
Hyades   & $2.86 \pm 0.03$ & $0.71 \pm  0.01$ & $0.20 \pm 0.02$ & $0.15 \pm 0.01$ & $11.07 \pm 1.01$ & $0.56 \pm 0.05$ & $0.09 \pm 0.01$  \\
M37      & $3.10 \pm 0.07$ & $0.80 \pm  0.03$ & $0.23 \pm 0.04$ & $0.18 \pm 0.03$ & $13.75 \pm 2.21$ & $0.71 \pm 0.12$ & $0.06 \pm 0.02$  \\
Praesepe & $3.21 \pm 0.16$ & $0.79 \pm  0.02$ & $0.20 \pm 0.05$ & $0.15 \pm 0.04$ & $11.70 \pm 2.74$ & $0.62 \pm 0.14$ & $0.07 \pm 0.03$  \\
NGC 3532 & $3.57 \pm 0.21$ & $0.85 \pm  0.02$ & $0.16 \pm 0.05$ & $0.13 \pm 0.05$ & $10.07 \pm 3.40$ & $0.56 \pm 0.19$ & $0.08 \pm 0.04$  \\
NGC 1039 & $4.00 \pm 0.23$ & $0.87 \pm  0.04$ & $0.09 \pm 0.06$ & $0.08 \pm 0.05$ & $5.97 \pm 3.85$ & $0.41 \pm 0.27$ & $0.09 \pm 0.04$  \\
M35      & $5.14 \pm 0.31$ & $0.96 \pm  0.03$ & $0.10 \pm 0.04$ & $0.09 \pm 0.03$ & $6.42 \pm 2.42$ & $0.70 \pm 0.43$ & $0.03 \pm 0.04$  \\
NGC 2516 & $5.47 \pm 0.59$ & $0.99 \pm  0.03$ & $0.13 \pm 0.06$ & $0.12 \pm 0.06$ & $8.59 \pm 4.37$ & $0.96 \pm 0.66$ & $0.00 \pm 0.06$  \\
\hline
Weighted mean & $2.65 \pm 0.02$ &  $0.73 \pm 0.01$ & $0.16 \pm 0.01$ & $0.12 \pm 0.01$ & $8.37  \pm 0.60$ & $0.56 \pm 0.04$ & $0.08 \pm 0.01$  \\
\enddata

\tablecomments{\label{tab:fuel} Initial mass, final mass, fractional core growth and corresponding energy output, and estimated helium yields of each cluster. (1): Cluster. (2): Progenitor Mass in \msol. (3): Remnant mass in \msol. (4): Fraction of remnant mass added to the core in the TP-AGB phase. (5): Core mass added in the TP-AGB phase in \msol. (6): Minimum total energy output required by TP-AGB core growth in $10^9\ L_{\odot} yr$. (7): Fraction of energy in TP-AGB phase predicted by \citet{Bertelli08, Bertelli09} accounted for by core growth. (8): Assuming the total energy output of TP-AGB stars in \citet{Bertelli08, Bertelli09}, estimated stellar yields of helium in \msol.}

\end{deluxetable}

\end{document}